\documentclass[11pt]{article}

\usepackage{graphicx}
\usepackage{latexsym}

\begin{document}
\title{On Relativistic Generalization of Gravitational Force}
\author{Anatoli Andrei Vankov\\         
{\small \it IPPE, Obninsk, Russia; Bethany College, KS;  anatolivankov@hotmail.com}}

\date{}

\maketitle

\begin{abstract}

The problem of point particle in the $1/r$ gravitational field was studied in SR-based Mechanics. Equations of motion under assumption of field dependent proper mass  were obtained in the relativistic Lagrangean framework. The dependence of proper mass on field strength was derived from the equations of particle motion. The result was the  elimination of $1/r$ divergence. It was shown that a photon in a gravitational field may be described in terms of a refracting massless medium. This makes the gravity phenomenon compatible with SR. New results concerning gravitational properties of particle and photon are discussed. The conclusion is made that the approach of field-dependent proper mass is perspective for further studies on divergence-free gravitational field development.


\end{abstract}

\section{Introduction}

The central question of this work is the one of Special Relativity mechanics: is a gravitational force compatible with SR? When investigating it, we do not use arguments from a quantum field theory, and do not question General Relativity. The objective of this work is to show how a gravitational force can be included in SR Mechanics framework. Though we use the term ``field'', it has a classical mechanics meaning of $1/r$ potential field, or the corresponding Minkowski force field. Occasionally we refer to some comparable results of ``conventional theories'' as far as it concerns problems of particle motion discussed in conventional SR Mechanics as well as in GR or classical field theories {\it under assumption of proper mass constancy}. A novelty of our approach is that the proper mass varies under the force action, and its dependence on field strength is found from the equations of motion.

At some historical stage of GR development, there were numerous attempts to incorporate the Newton's formulation of the gravitational law into SR as a starting point to a field theory development. A Newtonian field propagates with infinite velocity, and one could expect that this assumption would be automatically corrected in the covariant formulation of the gravitational law. Approaches were based on the concept of proper mass constancy and the concept of a photon coupling to the gravitational field: the latter was thought a necessary condition for explaining the observed bending of light (see \cite{Misner} and elsewhere). Not surprisingly, the attempts failed, first of all, because the stress-energy tensor of the electromagnetic field has a vanishing trace. Thus, SR Mechanics of a point particle under gravitational force action has never been developed.

We revisited this problem in the SR framework and studied the role of the proper mass in SR dynamics. The conclusion was made that the commonly used concept of the proper mass constancy is neither required in theory physical foundations nor it is justified by observations: so far, this is an arbitrary assumption, subject to theoretical examination and experimental verification. 

In our SR-based methodology of a variable proper mass, the world line is curved, while the metric remains the Minkowskian one: diagonal elements are functions of dynamical variables, off-diagonal terms identically equal zero. The equations of particle motion were derived, solutions to which {\it under weak-field conditions} were found similar to those in conventional theories (GR and its modifications). New results were predicted concerning gravitational properties of a particle {\it under strong-field conditions}. As for photon, the conclusion was made that it can be treated in a relativistic model, in which a field acts on the photon as an optically active medium. In other words, this is the gravitational refraction rather then force attraction that causes the bending of light. Thus, the issue of SR incompatibility with the gravity phenomenon took a new turn: the inclusion of gravitational forces into SR was justified.

It is shown that the concept of {\it variable} proper mass leads to a Lagrangean conservation symmetry and an  elimination of the $1/r$ divergence through the mechanism of proper mass ``exhaustion''. This is a new important result, which needs to be further investigated. Our idea of singularity elimination was presented earlier in (\cite{Vankov1}), and here we study different aspects of it in more details.

\section{Lagrangean Formulation of Relativistic Mechanics of Point Particle in Gravitational Field}

\subsection{Variable proper mass concept}
 
The following definitions and denotations are used. In Minkowski space of metric $\eta_{\mu \nu}$, any 4-vector $x^\mu$\ ($\mu=0, 1, 2 ,3$) is characterized by a time (temporal) component $x^0$ and a space (spatial) part, that is the 3-vector $x^i$ ($i=1, 2, 3$) in 3-space. The inner scalar product is defined ${\bf x\cdot x}=\eta_{\mu \nu}x^\mu x^\nu=x^\mu x_\mu=(x^0)^2-\sum_i (x^i)^2$. The position vector in the 4-coordinate Minkowski space is $x^\mu =(c_0t, x^i)$; the vector traces the trajectory of motion (the world line), which is not a straight line if a field is present. The corresponding proper velocity vector, which is tangent unit one, is a generalization of 3-velocity $v^i$: $u^\mu=dx^\mu/ds$, where $ds=\sqrt{ds^\mu ds_\mu}=c_0d\tau$ is the arc length interval of world line $s$, $c_0$ is the speed of light in the absence of field, and $d\tau=\gamma t$ is related to the so-called coordinate time interval $t$ in the formula $v^i=dx^i/dt$.

The 4-momentum vector is introduced as a generalization of the 3-momentum: $P^\mu=mu^\mu=m(dx^\mu/ds)$ with the obvious connection to $v^i$: $P^\mu=(m\gamma, \  m\gamma v^i/c_0)$. From this, the 4-momentum magnitude equals the proper mass $\sqrt{P^\mu P_\mu}=m$. 

An important stage in Relativistic Mechanics is the introduction of Minkowski force $K^\mu$ (so far not specified) acting on a test particle of the proper mass $m$. In GR and conventional Relativistic Mechanics the proper mass is assumed to be constant $m=m_0$, so the dynamics equation has the form 
\begin{equation} 
dP^{\mu}/ds=d(m_0 u^\mu) /ds=m_0 du^\mu /ds= K^\mu 
\label{l}
\end{equation}
We change the above assumption and consider the proper mass being field-dependent $m=m(s)$ to allow for a non-zero tangent Minkowski force component $u^{\mu}(dm/ds)$
\begin{equation} 
K^\mu=dP^\mu/ds=u^{\mu}(dm/ds)+m(du^{\mu}/ds) 
\label{2}
\end{equation}

The question arises: how does one know whether the proper mass is constant (as mostly assumed in current field theories) or field dependent (as suggested in this work)? Our viewpoint is that the proper mass constancy assumption is the issue of theory physical foundations and subject to experimental falsification. It should be noted that the proper mass variability is not a new idea: it was discussed in classical books on relativity theory by Synge \cite {Synge} and Moller \cite {Moller} and occasionally later on in connections with field theories but did not draw much attention among physics community. We are going to confirm that the introduction of the field-dependent proper mass in the relativistic Lagrangean framework leads to a consistent relativistic mechanics.

\subsection{A relativistic generalization of static gravitational force}

Consider a test particle characterized by a field-dependent proper mass $m$. Let the particle be slowly moved at a constant speed along the radial direction in the $1/r$ static gravitational potential field due to a spherical source of a radius $R$ and a mass $M_0>> m_0$, where $m_0$ is a particle proper mass at infinity. Such an imaginary experiment can be done by means of an ideal
transporting device provided with a recuperating battery. Work on the particle of a variable proper mass is given by:
\begin{equation} 
F(r)dr=m(r)c_0^2 d(r_g/r), \  \  \ (r\ge R)
\label{3}
\end{equation}
where $r_g=GM_0/c_0^2$ is a gravitational interaction radius.
Since the gravitational force is compensated by a reaction from the transporting device, the particle must exchange energy with the battery in a process of
mass-energy transformation. So the change of potential energy is related to the
proper mass change:
\begin{equation}
dm(r)=-m(r)d(r_g/r), \ \ r\ge R
\label{4}
\end{equation}
and the proper mass of the particle is a function of $r$:
\begin{equation}
m(r)=m_0\exp(-r_g/r), \ \ r\ge R
\label{5}
\end{equation}
In a weak field approximation $r\ge R>>r_g$, we have a Newtonian limit, and still can retain the proper mass variation:
\begin{equation}
m(r)\cong m_0(1-r_g/r),
\label{6}
\end{equation}
As is seen from (\ref{5}), the proper mass tends to exhaust as $(r_g/r)$ rises,       
while a gravitational potential energy takes the form:
\begin{equation}
W(r)=-m_0c^2[1-\exp (-r_g/r)]
\label{7}
\end{equation}
and the force work is given by
\begin{equation}
F(r)dr=d W(r)=m_0c_0^2{\cdot} \exp (-r_g/r)d(r_g/r)
\label{8}
\end{equation}
The potential energy changes within the range $-m_0c^2\le W(r)\le 0$. Therefore, it
is limited by the factor $c^2$, and a divergence of gravitational energy is naturally eliminated. The same will be shown true for a particle in free fall. 

It is interesting to note that in the time of GR development, Finnish physicist G. Nordstroem \cite{Nordstroem} tried to develop an alternative gravitational mechanics and field theory. Obviously, he was aware of option (\ref{2}), in which the proper mass depends on a gravitational potential $\phi(r)$. In 1912-13 he considered a formulae $m(r)=m_0\exp (-g\phi)$ with some ``adjusting factor'' $g$. Having troubles with gravitational properties of light and inertial mass, he did not come to a consistent theory and abandoned work after Einstein's GR was published in 1915.  

From  (\ref{5}) it follows  that a predicte deviation from $1/r$ potential is noticeable near a source of high mss density, and it is not realistic yet to observe the effect in laboratories. Nevertheless, challenging experiments are in progress. In one of them, an alleged test of a supersymmetry theory prediction of the $1/r^2$ law violation is attempted with the use of a symmetric torsion pendulum  \cite{Hoyle}. The authors look for a quite large correction $[1+\alpha \exp{(-r/\lambda)]}$ in a direction, which is opposite to what we predict. Their assessment of the effect was obtained by conventional mechanics methods based on the gravitational force concept, in fact, similar to that of mechanical force: the kinetic energy gain $(\gamma-1)m_0$ is taken from an ``inexhaustible''  source. For this reason, potential energy is subject to $1/r$ divergence. We are motivated by the prediction of a new phenomenon, the proper mass exhaustion (\ref{5}) under strong field conditions. The phenomenon leads to a natural elimination of the divergence.

\subsection{Relativistic Lagrangean formulation of the problem}

For a particle of variable proper mass $m(s)$, $s=s({x_\mu})$, in a gravitational $1/r$ potential field, it is convenient to introduce {\it a proper Lagrangian} $L(s)$ in order to exploit the Minkowski force concept $K^\mu = -\partial{W}/\partial{x_\mu}$. (In fact, we cannot formulate the Lagrangian in terms of coordinate time $t$ since a relationship of Minkowski and ``ordinary'' forces is not known prior to proper Lagrangian study).   
A relativistic analog to the difference of kinetic and potential energy in our case is $(m_0-m)$ with $m u^\mu u_\mu=P^\mu u_\mu $. Because of the identity $u^\mu u_\mu=1$ and the source being stationary, the Lagrangian should not an explicit function of $u^\mu$ or $s$. The proper mass must monotonously decrease as the particle approaches the source since a mass defect is associated with a growing binding energy. Yet, the potential field concept requires that $m\to m_0$, $W\to 0$ at infinity. In terms of the Noether's theorem, the $s$-translation symmetry, or the $t$-translation in $(t, x^i)$ coordinate system, is a manifestation of relativistic total energy conservation of a particle in a potential field.

We are going to study the time translation symmetry in Euler-Lagrange equations derived from the Hamilton's action principle. A relationship between any type of symmetry of a dynamical system with a conservation of corresponding quantity (Noether's current) is elegantly follows from the famous Noether's theorem. Her method works in a spirit of Hamilton's reformulation of Lagrangean mechanics. A trivial example is a classical system characterized by a set of generalized dynamical variables 
$[q(t),\ \dot q (t)]$ and a Lagrangian $L[(q(t),\ \dot q (t)]$, a system evolution is determined by the Euler-Lagrange equations  
\begin{equation}
\frac{d}{dt}[{\partial L}/{\partial \dot q}]={\partial L}/{\partial q}
\label{9}
\end{equation}
If the r.h.s. of (\ref{9}) is zero (the system has a $q$-symmetry), a quantity ${\partial L}/{\partial \dot q}$ is conserved. If the system additionally has the time symmetry, then 
$d L(q, \dot q)]/dt-[{\partial L }/{\partial q)}{\dot q }+ ({\partial L }/{\partial \dot q})({\partial  \dot q) }/{\partial t) })] = 0$. From this, in combination with (\ref{9}), the conserved Noether's current 
$ j=[\dot q ({\partial L }/{\partial \dot q}) - L]$ is derived. It characterizes a sum of kinetic and potential 
energy, the Hamiltonian $H=(T+W)$. 

Back to our problem: having the term $T=P^\mu u_\mu=m$ in the proper Lagrangian, one gets the Noether's conserved current $j=m_0-m+W=0$ (a change of kinetic energy equals a change of potential energy, their sum equals zero). It satisfies the requirement $(m_0-m)\to 0$, $W\to 0$ at infinity ($W\le 0$).
With the inclusion of rest mass, the conserved current is total energy, the Hamiltonian
\begin{equation}
H=m_0+T+W=m_0
\label{10}
\end{equation}
We shall return to this issue later in discussions of relativistic Euler-Lagrange equations.

\subsection{Equations of motion}

As was explained, the stationary Lagrangian is given by  
\begin{equation} 
L(s)=-m(s)-W(s)
\label{11}
\end{equation}
where $s=s(x^\mu)$ is a world line (arc)length, and a field is characterized by potential energy $W(s)$ (it is  negative for an attractive force). 
The Euler-Lagrange equations of motion follow from Hamilton's principle of the extremal action $S$
\begin{equation} 
\delta S=\delta \int_a^b L(s)ds=\int_a^b (\delta L) ds + \int_a^b  L d(\delta s)= \delta S_1 + \delta S_2 = 0 
\label{12}
\end{equation}
with a set of dynamical variables $x^\mu$ (the $s$ is not the one). Obviously, the proper mass $m(s)$ should not be considered an additional dynamical variable in a sense of the fifth degree of freedom. Thereafter, $m(s)$, 
$W(s)$, u(s), $s$, and $ds$ are subject to variation through independent variations of $x^\mu$. The proper velocity $u^\mu(s)$ as a function of dynamic variables $x^\mu$ will appear in the variational procedure, as well. 

It should be noted that the relativistic Lagrangean problem for a free particle motion was discussed in \cite{Landau},  \cite{Goldstein} with $W(s)=0$, the 
Lagrangian $L(s)=-m_0$ (in our denotations) and the action variation 
$\delta S=m_0 \delta \int_a^b ds=m_0\int_a^b d(\delta s)=0$. Clearly, this is a particular case of (\ref{12}).

From (\ref{12}) to continue, we have
\begin{equation} 
\delta S_1= \int_a^b (\delta L) ds=\int_a^b \frac{\partial L(s)}{\partial s}u^\mu\delta x_\mu ds                        
\label{13}
\end{equation}
\begin{equation} 
\delta S_2= \int_a^b L \delta(ds)=\int_a^b L \delta (u^\mu) dx_\mu=\int_a^b  L \frac{\partial u^\mu}{\partial s} \delta x_\mu ds                        
\label{14}
\end{equation}
\begin{equation} 
\delta S=\delta S_1+\delta S_2= \int_a^b \frac{d}{ds} (L u^\mu) \delta x_\mu ds=0                        
\label{15}
\end{equation}
Because variations $\delta x_\mu$  between the end points are independent for different $\mu$, the equality $\delta S=0$ in (\ref{15}) is possible if and only if 
\begin{equation}
\frac{d \left[L(s)u^\mu(s)\right]}{d s} =0
\label{16}
\end{equation}
With the Lagrangian (\ref{11}) substituted into (\ref{15}), we have Euler-Lagrange equations of motion
\begin{equation}
\frac{\partial \left[m(s)u^\mu(s)\right]}{\partial s}=-\frac{\partial \left[W(s)u^\mu(s)\right]}{\partial s}
\label{17}
\end{equation}
Having the additional equation of time-likeness of particle motion
\begin{equation}
 u^\mu u_\mu=1, \  \ u^\mu (du_\mu/ds)=0
\label{18}
\end{equation}
one is able to determine five correlated quantities $x^\mu(s)$, $m(s)$. 
Finally, one needs to introduce Minkowski force $K^\mu=-u^\mu \left ( {\partial W}/ {\partial s}\right)$ to get the desired equation of motion in terms of 4-momentum rate and Minkowski force
\begin{equation}
\frac{d}{ds}\left(m u^\mu \right) = K^\mu
\label{19}
\end{equation}

There are, actually, two orthogonal (vector) equations in (\ref{19})  
\begin{eqnarray} 
u^\mu (dm/ds)=K^\mu_{tan}, & \  m (du^\mu/ds)=K^\mu_{per}
\label{20}
\end{eqnarray}
where  $ u^\mu (dm/ds)=K^\mu_{tan}= -u^\mu {\partial W}/{\partial s}$ is a tangential component, and  
$m (du^\mu/ds)= K_{per}^\mu =-W (du^\mu/ds)$ is due to a Minkowski force component acting perpendicularly to the world line. The two equations are coupled in a feedback manner through a varying proper mass. From the scalar product $P^\mu u_\mu$ and (\ref{18}), the following useful formulae are obtained: 
\begin{eqnarray} 
 K^\mu u_\mu = dm/ds, &  K^0 u_0 \ = dm/ds + \ K^i u_i & (i=1,\ 2,\ 3) 
\label{21}
\end{eqnarray}
which express an energy balance (a current in 4-space). The existence of two orthogonal solutions is a consequence of proper mass variability under force action. This is a new result, significance of which is seen in applications.

\section{The ${\bf 1/r}$ Gravitational Potential}

\subsection{Equations of motion}

For the practical use of results obtained in previous sections, one should express (\ref{19}) in terms of time-dependent 3-space coordinates $x^i(t)$ using a connection of proper/improper quantities $ds=c_0 dt/\gamma$ and the definition of $P^\mu$. The $t$ is a ``wristwatch'' time measured by an observer at rest with respect to the source but far away from it (ideally, at infinity), as discussed later.

The spatial part of (\ref{19}) is given by 
\begin{equation}
\frac{d}{dt}(\gamma mv^i)=F^i 
\label{22}
\end{equation}
with the relationship between Minkowski and ordinary forces acting on a test particle in 3-space
\begin{equation}
F^i={\frac{c_0^2}{\gamma}}K^i
\label{23}
\end{equation}
The second independent equation follows from the temporal part of (\ref{19}):
\begin{equation}
\frac{d}{dt}(\gamma m)=\frac{c_0}{\gamma}K^0
\label{24}
\end{equation}
which expresses the total energy rate of the particle in the field.  By definition of a conservative field, 
$K_0$, being a total energy rate, must be zero, hence, $\gamma m=C$. For the particle starting free fall from rest at infinity, $C=m_0$, $\gamma m=m_0$. This result will be later substantiated by considering the Noether's conservative current, which is recognized in (\ref{21}) or, equivalently
\begin{equation}
\frac{d}{dt}(\gamma m c_0^2)=F^i v_i+\frac{c_0^2}{\gamma}\frac{dm}{dt}
\label{25}
\end{equation}

Further we are to restrain ourself to the problem of free radial fall; an orbital motion is subject to a separate work. Thus, $dr(t)=c_0\beta(t) dt$, and (\ref{25}) becomes
\begin{equation}
\gamma d(\gamma m c_0^2)=\gamma F(r)dr+c_0^2 dm
\label{26}
\end{equation}
which is the total energy balance in a differential form. In fact, this is the Noether's conservative current discussed earlier in terms of proper quantities and now expressed in the  ($r,\ t$) coordinates in the differential form. It manifests a total energy conservation law for a particle in a spherical symmetric potential field: ``the conserved total energy'' equals a sum of ``the potential energy change due to gravitational force work'' and the corresponding  ``kinetic energy change'', where the total energy is $\gamma m=m_0$ in the considered case of free fall from rest at infinity. Therefore, the l.h.s. of (\ref{26}) is zero. 

Next step is to substitute the gravitational force expression (\ref{3}) into (\ref{22}) (or equivalently (\ref{26})) to find the proper mass function $m(r)$ taking into account the conservation $\gamma m=m_0$. With a denotation $\gamma_r=m_0/m(r)$, the equation for radial motion takes the form $m_0c_0^2 \gamma\beta d\beta=m_0c_0^2 d(r_g/r)$ with the dynamical solution to it
\begin{equation}
1/\gamma_r=m(r)/m_0=1-r_g/r, \  \ m(r)=m_0(1-r_g/r)
\label{27}
\end{equation}
where $r=r(t)$,  $\gamma_r (t)=\gamma[r(t)]$ that is, $\gamma m=m_0$ with $m[r(t)]$ as a function of $r$ in 
(\ref{27}).
It looks like a linear approximation (\ref{6}) of the static relation (\ref{5}) and consequently has a range restriction
$(r\ge R> r_g)$, discussed later. 
From this solution, kinetic energy as a difference of total and proper energy is 
\begin{eqnarray}
E_{kin}=m_0 c_0^2-m(r) c_0^2 =m c_0^2 (r_g/r) & \ (r\ge R> r_g)
\label{28}
\end{eqnarray}
while the sum of kinetic and potential energy equals zero what makes the total energy $E_{tot}=m_0 c_0^2$. By finding the specific function $m(r)$ (\ref{27}), we confirmed the Noether's current concept (\ref{10}) and the constancy $\gamma m=m_0$.

If the particle in radial fall has kinetic energy at infinity $E_0=\gamma_0 m_0^2$ then, due to the total energy conservation, $m_0$ should be replaced by $\gamma_0 m_0$; correspondingly, the equality $\gamma(t)=\gamma_r(t)$ should be replaced by $\gamma (t)=\gamma_0 \gamma_r (t)$, where $\gamma_r=m_0/m(r)$, $r=r(t)$ as before, $\gamma_0=(1-v_0^2/c_0^2)^{-1/2}$, $v_0$ is the radial speed at infinity. Then (\ref{28}) becomes 
\begin{eqnarray}
E_{kin}= m c_0^2(\gamma_0-1)+r_g/r
\label{29}
\end{eqnarray}
However, this is not a final result because we need to take into account the mass defect in the gravitational force expression (considered next).

\subsection{Correction for the source mass defect, and final results.}

Our requirement of $(r_g<R)$ in (\ref{28}) precludes the proper mass from reaching a zero value in the exterior region when $m\to 0$ at $r\to r_g$. The problem is caused by the simplified concept of the gravitational radius $r_g=GM/ c_0^2$, in which a binding energy of the sphere (a mass defect) was ignored because the interior solution for $r<r_g$ was not studied. We need to take into account the fact that $M\ne  M_0=\sum_i {m_{0i}}$, where $m_{i0}$ are proper masses ``at infinity'' of particles comprising the sphere. The difference is a self-binding energy $\Delta M=M_0-M$. One needs to reformulate the problem in terms of $r_{g0}=GM_0 / c_0^2$ with the correction for the mass defect. An approximate way to do it would be to introduce a spacial factor
$M_0/M=m_0/m=\gamma_r (r)$. Then, the gravitational force takes the form  
\begin{eqnarray}
F(r)dr=GM_0 (m^2/m_0)dr(1/r)=m_0c_0^2(m/m_0)^2 d(r_{g0}/r) 
\label{30}
\end{eqnarray}
The correction ensures physical requirement $(m(r)>0)$ in the whole range $(r>R)$ and a boundary junction of exterior solution $m(r)$ at $(r\ge R)$ with that at the surface $r=R$ without actual finding the interior solution. Further on, we drop the lower zero index in $r_{g0}$ and use the previous denotation $r_g=GM_0/c_0^2$ for the gravitational radius having a new meaning. The introduction of the additional factor $\gamma_r=m_0/m$ in the source term is an approximate way to account for the source self-binding effect in order to correct a radial dependence of an exterior field under strong field conditions. 

All things considered, the equation (\ref{22}) takes the form 
\begin{eqnarray}
\gamma^2 \beta d\beta=d(r_{g}/r) 
\label{31}
\end{eqnarray}
and the dynamical solution is: 
\begin{eqnarray}
1/\gamma_r= m/m_0=\exp{(-r_{g}/r)}, \ &  (r\ge R) 
\label{32}
\end{eqnarray}
It coincides with the static solution  (\ref{5}). Having kinetic energy term $\gamma_0$ been accounted for from the condition at infinity,
we have a final set of formulae:
\begin{eqnarray}
\gamma=\gamma_0\gamma_r=\gamma_0\exp(r_g/r), & 
\beta(r)=\left[1-(1/\gamma_0^2)\exp(-2r_g/r)\right]^{1/2} 
\label{33}
\end{eqnarray}
and squared norms of the 4-momentum $P^\mu=m(\gamma,\gamma\beta,0,0)$  and the 4-coordinate vector 
$\Delta x^\mu=c_0\Delta\tau(\gamma,\gamma\beta,0,0)$
\begin{eqnarray}
(c_0 m)^2=(c_0 \gamma m)^2- p^2  
\label{34}
\end{eqnarray}
\begin{eqnarray}
 (\Delta s)^2=(c_0 \gamma \Delta \tau)^2- (\Delta r)^2  
\label{35}
\end{eqnarray}
Relations will be used further: $\gamma=\gamma_r \gamma_0$, $\gamma m=\gamma_0 m_0$, 
$\gamma \Delta\tau=\gamma_0 \Delta t_0$, $p=c_0 \gamma \beta m=c_0 \gamma_0 \beta m_0$, 
$\Delta s=c_0 \Delta\tau$,  
$\Delta r= c_0 \gamma\beta\Delta\tau= c_0 \gamma_0 \beta\Delta t_0$ ($t_0$ is the ``coordinate'' time measured by the rest observer at infinity; it is usually denoted $t$, as discussed later). 
 
Formulae for total, kinetic, and potential energy are:
\begin{equation}
(E_{tot}/c_0)^2=(\gamma_0 m_0 c_0)^2 =p^2+ (m c_0)^2 
\label{36}
\end{equation}
\begin{equation}
E_{kin}(r)=E_{tot}-m c_0^2=m_0 c_0^2 \left[\gamma_0-\exp(-r_g/r)\right] 
\label{37}
\end{equation}
\begin{eqnarray}
W(r)=-m_0 c_0^2\left[ 1-\exp{(-r_g/r)} \right] \ , &   (r\ge R)
\label{38} 
\end{eqnarray}
It is seen that \  $-m_0c_0^2 \le W \le 0$. When $m_0<<M_0$, the kinetic energy emerges solely due to the change of the proper mass of a test particle in a field, and the proper mass ``exhaustion'' under strong field conditions takes place. We want to emphasize again that
the divergence is eliminated for an arbitrary mass density of the source and a however strong field.

Under weak-field conditions $r_g/r<<1$, we have
\begin{equation}
\gamma=\gamma_0(1+r_g/r), \ \ \beta=\left[1-(1-r_g/r)/\gamma_0) \right]^{1/2} 
\label{39}
\end{equation}
\begin{equation}
E_{kin}=m_0 c_0^2(\gamma_0-1+r_g/r) 
\label{40}
\end{equation}
\begin{equation}
W(r)=-m_0 c_0^2 (r_g/r),  \  \phi(r)=W(r)/m_0 c_0^2=-(r_g/r)  
\label{41}
\end{equation}
and the Newtonian limit $E_{kin}=mv^2 /2$.

Clearly, our results and conventional ones differ due to the difference in concepts of relativistic mass and, correspondingly, potential energy. A particle to be accelerated by a force at distance needs to be bound. The binding energy in our philosophy is a real mass defect $(m-m_0)$ limited by the proper mass value. It makes the force weaken as $r\to r_g$ so that no infinities arise. In the concept of proper mass constancy, the particle gets bound while acquiring kinetic energy from field energy $m_0c_0^2 (r_g/r)$, so both the binding and kinetic energy, in principle, are unlimited.

\section{Lagrangian symmetry, Noether's theorem, and energy conservation}

\subsection{Time-translation symmetry, and energy conservation}

In order to study the Noether's current in more details, let us go back to (\ref{34}), (\ref{35}) to consider a world line in the 4-momentum space in a manner as we do in the 4-coordinate space, and compare 4-vector norms:
$ \Delta S(r)=|\Delta {\bf x|}$ and $ \Delta S_p (r)=|\Delta {\bf P|}$ of the coordinate vector 
$\Delta x^\mu=c_0 \Delta t (1,\ \beta,\ 0,\ 0)$ and the momentum one $c_0 P^\mu= c_0 m_0 (1,\ \beta,\ 0,\ 0)$, respectively. In the case of a radial motion from rest at infinity, the Lorentzian norms are:   
\begin{equation} 
\Delta S(r)=\left[(c_0 \Delta t_0)^2- (\Delta r)^2\right]^{1/2}=c_0\Delta t_0/\gamma_r=c_0\Delta\tau(r) 
\label{42}
\end{equation}
\begin{equation}
\Delta S_p (r)=\left[(c_0 m_0)^2- (p(r))^2\right]^{1/2}=c_0 m_0/\gamma_r=m(r) 
\label{43}
\end{equation}
where $r=r(t)$,  $c_0 \beta(r)=\Delta r/\Delta t=\Delta r'/\Delta\tau(r)$. The operational meaning is, as next. $\Delta r=c_0\beta \Delta t_0 $ is measured by a ``far-away observer'' at rest with respect to the source. She determines $\beta$ from measured $\Delta r$ per a constant time interval $\Delta t_0$ by the time-of-flight technique with the use of standard clocks and rods. Thus, we term $t=t_0$ with zero subscript ``a far-away time'', also called ``a coordinate time''. Next quantities are the contracted radial interval  $\Delta r'=\Delta r/\gamma$, and the world line interval $\Delta s(r)=c_0\Delta\tau(r)$ both measured by a comoving observer. The contraction is a pure SR kinematical effect. It is seen that  $\Delta s(r)=c_0 \Delta \tau(r)$ is not invariant.
 Notice that $\beta=\Delta r/\Delta t_0=\Delta r'/\Delta \tau$.

From measurements of the speed $\beta$, the gravitational time dilation effect can be obtained. The latter is associated with the proper mass dependence on the gravitational potential $m(r)=m_0\exp(-r_g/r)$. The corresponding frequency of atomic clock of the proper mass $m(r)$ is proportional to the proper mass: $m(r)c_0^2=h f(r)$, where $f(r)=1/T(r)$ is a relationship of $f(r)$  with the proper period 
$T(r)=\gamma_r \Delta t_0$ recorded by a local standard clock at rest at point $r$. Hence, 
$T(r)=\Delta t_0$ at infinity. The clock put at a deeper potential level $r_2\to r_1$, $r_2>r_1$ will slow down by the factor $\gamma_r$ in agreement with observations. Therefore, one needs the factor $\gamma_r=(1-\beta^2)^{1/2}=\exp{(r_g/r)}$ from measured values of $\beta$ for $\gamma_0=1$ to find $m(r)=m_0/\gamma_r$, $f(r)=f_0/\gamma_r$. There is a useful relationship $\Delta\tau (r) T(r)=\Delta t_0^2 $.  It becomes clear that the assumption of the proper mass constancy in the SR-based mechanics would result in a failure of a gravitational time dilation prediction. This is one of the reasons to discard the assumption of proper mass constancy in the SR framework. Changing the assumption makes a desired difference.

One can recognize a new conservation symmetry by examining 4-vector components in (\ref{42}), (\ref{43}) and rearranged in (\ref{44}); $\gamma_0$ is put equal to unit for simplicity there. Given conditions at infinity, conserved quantities are seen on the l.h.s. of each equation in (\ref{44}):
\begin{equation}
[c_0\Delta t_0]^2=[\Delta S (r)]^2 + [\Delta r]^2,\   \  
(c_0 m_0)^2 = [\Delta S_p (r)]^2 + [p(r)]^2  
\label{44}
\end{equation}
Instead of hyperbolic rotation in  metric (+, -, -, -), a real rotation symmetry emerged in the  4-vector representation in a quasi-Euclidean geometry of signature (+, +, +, +). The constant radius of rotation is 
$c_0 \Delta t_0$ and $c_0 m_0$ in coordinate and momentum space, correspondingly. The rotation angle $\theta$ is determined by $\sin\theta=\beta [r(t)] $,\ or identically \ $\cos\theta=1/\gamma=\gamma [r(t)]$. Compare it with an imaginary angle $\psi$ of hyperbolic rotation: $\cosh\psi=\gamma$, $\sinh\psi=\gamma\beta$, 
$\tan\psi=\beta$. Hence, $\sin\theta=\tanh\psi=\beta$.

\subsection{Total energy conservation law and dynamical complementarity principle}

The new (real rotation) symmetry ensures the total energy conservation law in the approach of the variable proper mass concept. It can be shown that a similar symmetry takes place under general conditions at infinity when $(\gamma_0>0)$ or  $(\gamma_0<0)$; the case of a negative initial kinetic energy at infinity means that the test particle is dropped at some finite point $r>R$ where the potential is not zero. The energy conservation is interpreted in terms of Nether's conserved mass-energy current in the momentum space. Our finding is that there is a similar conserved current in the coordinate space. It corresponds to the constant time rate recorded by a far-away atomic clock. Therefore, there are equivalent symmetries in $P^\mu$ and $x^\mu$ spaces.  This fact is known and used in the SR Kinematics, when the Klein-Gordon equation is derived in the SR Kinematics framework with the relativistic generalization of the de Broglie wave concept. The latter includes such quantities as the 4-phase $\phi=(\omega t -{\bf k\cdot r})$, the 4-wave vector $(E/c_0,\ {\bf p})=\hbar/c_0(\omega,\ {\bf k})$, where $E=m c_0^2=\hbar \omega=hf$; consequently, $P^\mu \sim f^\mu \sim K^\mu$ in SR Kinematics. The following scalar product is Lorentz invariant:
\begin{eqnarray}
K^\mu \Delta x_\mu= c_0P^\mu \Delta x_\mu=h \  \ \mbox{or} \  \ c_0P^\mu  x_\mu=N h 
\label{45}
\end{eqnarray}
where $N$ is a number of wavelength (clock ticks).
It is not surprising that the generalized de Broglie concept is valid in our {\it gravitational dynamical problem} in the quasi-Euclidean representation (\ref{44}), the invariance of the scalar product (\ref{45}) in the quasi-Euclidean metric takes place as well. The operational meaning of it is clear: all observers agree to use standard atomic clocks of the proper mass $m_0$ at infinity. The clock is considered a quantum oscillator in the de Broglie wave concept ($\Delta t_0$ and $m_0$ are reciprocal quantities) in the field-dependent proper mass approach. 

The fact of invariance (\ref{44}) in the quasi-Euclidean dynamical metric is called further  "the dynamical complementarity principle" due to its significance in our study. The  quantum de Broglie concept is seen to be naturally embedded in our SR-based gravitational dynamics before a field theory development. Some other issues relevant to the problem are discussed in \cite{Vankov2}. We believe that the real rotation symmetry is a true  gravitational mechanics law to be confirmed by observations; it reflects the idea of mass and time unity and enables us to gain into a new insight of physical and philosophical concepts of matter and time. 

\subsection{Graphical illustrations, and lessons}

A brief comment is needed before discussing graphical illustrations of the free fall problem. In SR textbooks, the Lorentz kinematical transformation is usually illustrated by a straightforward picture of hyperbolic rotation. This would be a trigonometrical rotation in a complex plane by an imaginary angle $\psi$, $\tan {\psi}=\imath {\beta}$; optionally, it can be shown as a hyperbolic rotation in a real plane so that $\tanh {\phi}=\beta$, hence, $\tan {\psi}=\tanh {\phi}$. The idea in both variants is to show in the graph the invariant Lorentz norm $\Delta s$ {\it as a rotating radius}. 

Our graphic presentation is different and has more physical sense for us. There are three terms depicted in each graph in a real plane: ``spatial part'' versus ``Lorentzian norm''. The picture presents the Lorentzian quadratic metric: ``squared Lorentzian norm'' = ``squared temporal part'' - ``squared spatial part'', and at the same time the quasi-Euclidean one: ``squared temporal part'' =  ``squared Lorentzian norm''+ ``squared spatial part''. In the second case, a ``temporal part'' (not the Lorentzian norm) rotates in a real plane by a real angle $\theta=tan^{-1} (\gamma\beta)$. It is possible now to illustrate the norm invariance in usual Lorentz-boost transformations (the case of inertial motion) as well as a real rotation in the case of free radial fall. In Fig.\ref{Sym} each graph is presented equivalently in $(p,\ m)$ and $(r,\ \tau)$ planes of $P^\mu$ and $x^\mu$ Minkowski spaces, correspondingly. 

Let us start with the case of the pure (no field) Minkowski space. There are three graphs $a)$, $b)$, and $c)$, which are different in a type of constraints imposed on Lorentz {\it kinematical} transformations in the 4-coordinate and 4-momentum spaces.



\begin{figure}[t]
\includegraphics{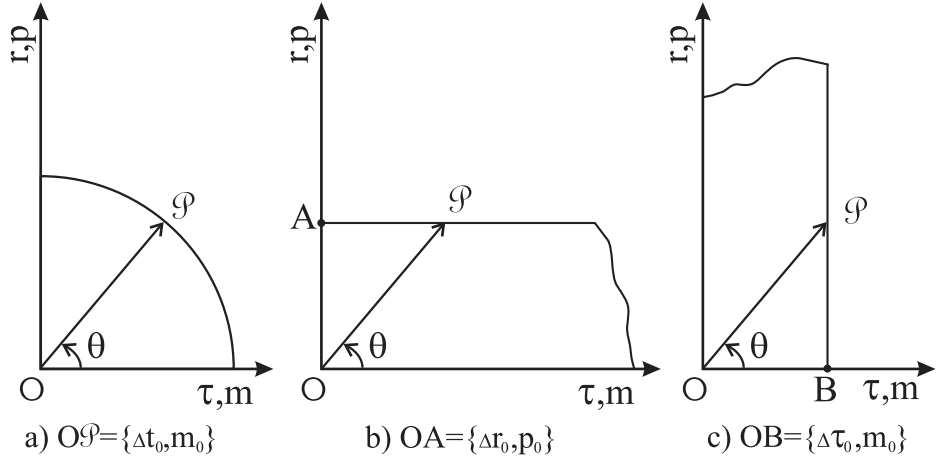}
\label{Sym}
\caption{\label{Sym}  4-vector kinematical and dynamical  rotation. $\theta=\tan^{-1}{(\gamma\beta)}$. 
{\it Graph a) SR Dynamics ($1/r$ potential field): real rotation symmetry. \  \  \  \  \ 
Graph b) Spatial part is fixed. There is no symmetries or invariance.\  \ \  \  \  \  \
Graph c) SR Kinematics (no field): proper mass and time Lorentz invariance.}}

\end{figure}

{\it Graph $c)$} presents a family of world lines with the parameter $\beta$ in pure (no field) Minkowski space. This is the case when observers travel with different speed provided the travel proper time $\Delta\tau_0$ measured by the traveling observer is fixed. ``The staying at home'' and traveling observers agreed to use standard atomic clocks to verify $\Delta\tau$ constancy. Consequently, both the proper time and the proper mass in the family of world lines are Lorentz invariant. Lorentz invariance does not takes place in other than {\it case c)} situations, as seen next.
   
{\it Graphs $b)$} describes the problem of travel with a speed $\beta$ (as a parameter) from $O$ to $A$ of a fixed distance, a constraint, $OA=\Delta r=\Delta r_0=c_0 \beta\Delta t$ in a pure Minkowski space. 
{\it Graph $a)$} is the case of the constraint $\Delta t=\Delta t_0$ (the travel time $\Delta t=\Delta t_0$ is fixed in the far-away observer's coordinate system). In both cases, it is not possible to make an agreement to use standard clocks: both the proper time and the proper mass depend on $\beta$ (they are not Lorentz invariant, and the kinematical complementarity does not hold).

Now, let us discuss our {\it SR-based dynamical problem} of free radial fall in the {\it graph $a)$}, $m(r)=m_0/\gamma_r$, $r=r(t)$. 
The graph illustrates the quasi-Euclidean representation of Minkowski 4-vectors. The family is produced in a single experiment with one freely falling particle, the world line of which is divided into small adjacent intervals (partitions), the Lorentzian norm $\Delta s =c_0 \Delta\tau $. Conditions at infinity are fixed: $\theta=0$, $\Delta\tau=\Delta t=\Delta t_0$; ($\gamma_0=1$ for simplicity). The dynamical complementarity principle and the time translation symmetry hold. The Lorentzian norm and the proper mass are functions of $\beta$, while the vector $OP$ is a conserved temporal component. The graph shows a real rotation of $OP$ with a constant radius of rotation (the Noether's current) 
$OP=c_0 \gamma\beta\Delta\tau=c_0 \beta\Delta t_0$ in $x^\mu$ space and 
$OP=\gamma m =m_0$ in $P^\mu$ space. The angle $\theta(\beta)$ is a function of dynamic variables $x^\mu$. It characterizes an instantaneous state of a freely falling particle at an instant $t$: $\theta (r)=\sin^{-1} \beta (r)$, $r=r(t)$. Obviously, {\it graphs $b)$} and {\it $c)$} are not relevant to the problem. 

Some lessons are drawn from graphs.

1. There are two categories of constant physical quantities. The first category relates to the hyperbolic rotation in pure Minkowski space. The constancy is due to constraints imposed on the Lorentzian vector representation. Only the constraint $c)$ leads to the kinematical complementarity principle and  Lorentz  invariance under $\beta$-boost transformations in 4-coordinate and 4-momentum complementary spaces. The Lorentz invariance is due to the translation symmetry of a 4-point in Minkowski $x^\mu$ and $P^\mu$ space at the same time. An attempt to construct an ``extended'' Lorentz group without respecting the complementarity principle would mean the abuse of the Minkowski space concept.


2. One should distinguish the first (kinematical) category of Lorentz invariant quantities from the second (dynamical) category of conserved (unchanged in time) quantities in a Lagrangian system, the Noether's theorem deal with (as in {\it case $a$}). Our object under investigation is a 4-vector undergoing an evolution in the Lagrangian dynamical system in the Minkowski space. One can think of $\Delta s$ as an ``instantaneous image'' of the proper 4-position vector $OP$ tracing a small linear interval $s\to (s+\Delta s)$ on the curved  world-line $s$ in the 4-coordinate space or $m\to (m+ \Delta m)$ in the 4-momentum space. In the picture $a)$, the interval is a $P$-projection on the $\tau$-axis, (or on the $m$-axis) and it is not constant: it gets smaller as the particle approaches the source. However, the interval $\Delta t=OP$ is preserved. To find the proportion $\Delta t_0/\Delta\tau(r)= T(r)/\Delta t_0$ and the time interval $T(r)$ referred to the gravitational time dilation, one needs to draw the tangent line at the point $P$ to the intersection with the horizontal axis.

3. The picture $a)$ illustrates SR-based gravitational dynamics, essential part of which is the field-dependent proper mass concept. The conservation takes the form of rotation symmetry in a real plane in a quasi-Euclidean 4-space, and it is associated with the Noether's conserved current due to the time translation symmetry, as in classical mechanics. 

Descriptions of free fall in space-time and in the 4-momentum space are formally identical. Indeed, the final equation of motion (\ref{31}) does not contain a mass of a test particle (as in classical mechanics). The parameter of physical importance is the gravitational interaction radius $r_g$ in the gauge factor $\gamma_r=r_g/r$: the factor determines Minkowski space deformation via space-time and mass-energy rescaling. On the one hand, the source  causes proper mass variation under Minkowski force action (the momentum space curvature). On the other hand, it makes the world line curved (the coordinate space curvature). Certainly, the two currents would follow from the Noether'e theorem, provided the complementarity principle was stated in the Lagrangean problem formulation.

4. ``Two currents'' means that in our originally formulated relativistic Lagrangian, the proper mass $m$ can be replaced by the complementary  quantity $\Delta\tau$ to allow the Minkowski force coming to the scene in the momentum $K_m^\mu$ representation (affecting the proper mass), or coordinate $K_{\tau}^\mu$ representation (affecting the proper time) with the equivalent outcome. To agree on this proposition, one has to think about acted by force particles in a broader concept of atomic clocks, or interacting quantum oscillators, probing both mass/energy and space/time local (generally evolving) metric in comparison with the constant background at infinity (this is simply a suggestion to consider the de Broglie wave propagation in a gravitational field). Consequently, we deal with a complementary (double) Lagrangean formulation of the problem. As a result, there are two complementary solutions: 
\begin{eqnarray}
m=m_0\exp{(-r_g/r)}, & \Delta\tau=\Delta t_0\exp{(-r_g/r)} 
\label{46}
\end{eqnarray}
obtained in Sections 2 and 3 without emphasizing the fact of a double formulation. To check if it is true, just put $m=m_0/\gamma$ and $u^\mu=(\gamma, \gamma\beta, 0, 0)$ for $K_m$ in (\ref{19}), and do the same with $\Delta\tau=\Delta\tau_0/\gamma$ for $K_{\tau}$.  

In the next Section, we discuss the photon problem in the Minkowski (deformed) space. Instead of GR ``curved space-time field'', the more appropriate in SR Mechanics term is used: ``gravitational refracting medium''.



\section{A photon in the gravitational field}

Unlike the particle, the photon does not have a proper mass; its total mass is solely a kinetic one. One has to look for conserved quantities in the photon metric taking into account the photon SR kinematics \cite{Vankov3}. Instead of detailed analysis, we simplify the problem by considering the photon emitter/detector at rest with the respect to the source and making use of the fact that any photon in flight in a gravitational field is characterized by the two conserved quantities: an energy (frequency) and an angular momentum (the latter is out of consideration here). 

Thus, we assert that the energy (frequency) of the photon emitted at any point does not change during its travel in a gravitational field.  From the concept of the atomic clock, it follows that the frequency $f_{ph}$ at the instant of emission must be proportional to the frequency of an atomic clock-emitter $f(r)=m(r)c_0^2/h=f_0\exp(-r_g/r)$, that is, the {\it emission} frequency is field dependent. Therefore, the momentum (or the wavelength) and the speed of light will proportionally change with respect to those values measured by the far-away observer in experiments with the standard photon from her clock-emitter. All said above is sufficient for the determination of photon gravitational properties in the model, in which Minkowski space filled with field is considered a transparent refracting medium.

The next set of formulae describe characteristics of the photon detected at a point $r$, if emitted at a point $r'$.
\begin{equation}
f_{ph}(r'\to r)=f_0\exp(-r_g/r')
\label{47}
\end{equation}
where $f_0$ is the photon frequency at infinity; the photon does not change the initial (emission) frequency during its flight. The photon speed (the speed of light) is
\begin{equation}
c_{ph}(r'\to r)=c_0\exp(-r_g/r)
\label{48}
\end{equation}
So far, we consider results valid for all frequencies (there is no dispersion); hence, a photon and light propagate similarly. The speed of light at detection point $r$ does not depend on a point of emission $r'$. Consequently, the photon wavelength is
\begin{equation}
\lambda_{ph}(r'\to r)=\lambda_0 \exp(r_g/r'-r_g/r)  
\label{49}
\end{equation}
It follows that the photon wavelength at any point of emission equals the wavelength at infinity $\lambda_0$. Finally, the proper period of a resonance line of atomic clock is
\begin{equation}
T_{res}(r')=1/f_{res}(r')=T_0 \exp(r_g/r')
\label{50}
\end{equation}
All quantities with ``zero'' subscript are measured at infinity. The speed of light is influenced by the gravitational potential according to (\ref{48}); further a dimensionless form is used 
\begin{equation}
\beta_{ph}(r)=c(r)/c_0=\exp(-r_g/r)
\label{51}
\end{equation}
This is the speed of light wave propagation. Physical processes described by the above formulae are time reversal in accordance with the energy conservation. Thus, the gravitational time dilation and the red shift are due to the field dependence of the emission frequency and the speed of light, provided the photon energy being conserved.

It is seen that the speed of light is constant on the equipotential surface $r=r_0$, and it may be termed a tangential, or arc speed. One can define also the radial (``coordinate'') speed $\tilde \beta_{ph}(r)$
\begin{equation}
\tilde \beta_{ph}(r)=\beta_{ph}(r)(dr/d\lambda)=  \exp(-2r_g/r)
\label{52}
\end{equation}
Under weak-field conditions, it coincides with the corresponding GR formula. 

We conclude that the photon propagates in a gravitational field as in a refracting medium with the index of gravitational  refraction $n_g=1/\tilde \beta_{ph}$. The refraction concept was discussed in the GR literature (see, for example  \cite{Moller}, \cite{Fock}, \cite{Fischbach}). It should be noted that there is no evidence that a static electric or magnetic field alone would affect the speed of the photon. 


\section{Predictions and Observations}

GR tests are related to weak-field conditions and usually presented in literature as a solid GR gravitodynamics confirmation of the curved space-time concept \cite{Clifford, Weinberg}. In fact, under those conditions of ``near-Newton'' limit, a behavior of a photon and atomic clock in our approach is similar to that in GR (in spite of different space-time philosophy). How well our approach fits all observations is a special issue; many details need to be further investigated. Here we are able to make only a brief review of basic facts.

\medskip
{\it 1. The gravitational red-shift and time dilation}

The term ``red-shift'' means that the wavelength of a photon emitted by an
atomic clock at some point of lower potential appears to be increased when
detected at some point of higher potential. Our interpretation of the red-shift was explained earlier: the effect is due to a combination of the gravitational shift of the emission-detection resonance line and the dependence of the speed of light on field strength while a frequency of a photon in flight being constant and equal to the emission frequency (47-50). The latter is proportional to the field dependent proper mass $f\sim m$ what causes the gravitational time dilation. This interpretation is consistent with total energy and angular momentum conservation laws in the field-dependent proper mass concept. 


\medskip
{\it 2. The bending of light}

The bending of light is due
to the ``gravitational refraction''. We conducted different calculations of the bending effect: 
using a refraction model, and using the angular momentum conservation; in both cases, the result was the same and similar to that in GR. 

\medskip
{\it 3. The time delay of light flight}

The time delay effect was measured in radar echo experiments with electromagnetic
pulses passing near the Sun. The effect can be calculated by 
integrating the time of light travel over the path with the field-dependent coordinate speed
(\ref{52}); the result will be equivalent to GR predictions. 

\medskip
{\it 4. Planetary perihelion precession and other astronomical observations}

This problem is related to a particle orbital motion in a gravitational field. It adds nothing new to our conclusion about absence of numerical difference in predictions of weak-field effects in the alternative versus conventional theory.  The perihelion precession can be assessed in GR by comparing radial and orbital frequencies in the Schwarzschild metric under weak-field approximation or in the post-Newtonian parameterization model. In the alternative approach, the corresponding physical treatment is equivalent to that in the effective potential model, in which dynamical quantities of orbital motion are influenced by the first-order field dependence of the proper mass in the Minkowski space.



\medskip
{\it 5. A particle in free fall in a gravitational field}

This is the case when we can compare predictions under high energy conditions. 
According to GR \cite{Misner}, a relative
speed of a particle in a radial fall is described by 
$\beta(r)=(1-2r_g/r)[1-(1-2r_g/r)/\gamma_0^2]^{1/2}$.
It shows that from the viewpoint of the observer 
at infinity a particle dropped from rest begins to accelerate, then at some point  
starts decelerating and eventually stops at $r=2r_g$. The bigger initial kinetic 
energy, the greater a "resisting" force arising so that the speed of the particle cannot exceed the 
coordinate speed of light. Strangely enough, if $\gamma_0 \ge\sqrt{3/2}$, 
the particle will never accelerate in a gravitational field, (see (\cite{Okun, Mashhoon}, and elsewhere). 

The GR formula should be compared with our result (\ref{39}):\  \
 $\beta(r)=\left[1-(1/\gamma_0^2)\exp(-2r_g/r)\right]^{1/2}$, which does not indicate any ``resisting force''.
 


\medskip
{\it 6. ``Black holes'' and other ``strong field'' observations}

There are astrophysical observations related to strong-field effects (the so-called black holes, radiating binary star systems, and others). Of course, there should be strong-field effects around astrophysical objects of super-high density. Practically, they might look like circumstantial evidence of ``black holes'' manifesting ``gravitational collapse'' and the corresponding ``light trap''. However, the idea of matter collapse into a singularity point in space seems to be an unnecessary ``new physics'' speculation. In our alternative approach, the gravitational time dilation could be however great; physical processes involving particle and photon motion in a strong field remain time-reversal and free of singularities. We predict an existence of extremely dense ordinary material formations of a strong gravitational pull without collapsing.

\section {Summary and Conclusion}

-  The problem of relativistic motion in a gravitational field was studied in the Special Relativity dynamics of point particle. The novelty of our approach to the problem is an introduction of the field dependent proper mass concept, as opposed to conventional assumption of the proper mass constancy. Historically, the SR-based gravitational dynamics has never been developed. It was believed that the gravity phenomenon and Special Relativity are incompatible. Gravitational properties of relativistic particles are also not easy to explain. General Relativity did explain the observed gravitational properties of particles and photons. As for to-day, GR has been thoroughly tested under weak-field conditions; however, strong-field effects still have not been verified in direct measurements. The long-standing, not thoroughly understood problem is the GR non-quantizibility. Another problem is associated with the strong-field $1/r$ divergence, which cannot be removed by a means of renormalization procedure. That is why alternative approaches to the gravitational problem are often speculated in literature.

-  We studied conservation properties of the $1/r$ gravitational potential in the relativistic Lagrange framework in the context of Noether's currents associated with the time and mass translation symmetry. A quantum connection of the theory via the generalized de Broglie wave concept was established and the complementarity principle in relativistic dynamics was formulated. The proper mass and time are scalars, which determine the temporal part of coordinate and momentum (complementary) 4-vectors characterizing the particle as a standard quantum oscillator, or a standard atomic clock. The complementarity principle requires that all observers use standard atomic clocks in the metric determination comparing to the time pace at infinity. The principle enables us to gain an insight into a unity of mass and time concepts and quantum  connections of relativistic gravitational dynamics due to the relationship $m_0 c_02 \Delta\tau_0= h$.

-  One of our findings is that a photon propagation may be described in terms of refraction in a gravitational field medium. This is not an unusual approach to the photon problem: a similar photon concept was from time to time considered in the GR framework. It means that the photon may not be energetically coupled to the gravitational field but be influenced by another (refraction) mechanism of gravitational interaction. We concluded that the inclusion of the photon refraction concept along with a revised proper mass concept into SR-based mechanics makes predictions consistent with existing gravitational (weak-field) observations. New predictions in strong field domain were made.

-  The photon does not have the proper mass. Consequently, it gives rise to the null Lorentzian metric. In SR methodology, the photon plays an important role in determination of both temporal and spatial parts of complementary 4-vectors by a means of information (photon) exchange between observers. One of our findings is that the relativistic  Lagrangean problem has a dual formulation in terms of complementary quantities. This makes the concept of a gravitational field as a refracting medium more understandable but still does not give a clue about the mechanism of changing the speed of light (permittivity and permeability of space) in the field. The question of ``refracting'' properties of a gravitational field must be challenging for quantum gravity researchers.

-  The motivation of this work is new results in a strong field domain, in particularly, the $1/r$ divergence elimination through a natural mechanism of mass defect rising with field strength (the predicted ``mass exhaustion'' effect). We believe that the approach developed in the SR-based framework will be perspective for further studies on developing a divergence-free gravitational field theory.

\end{document}